\documentclass[%
 aip,
 amsmath,amssymb,
 reprint,%
]{revtex4-1}

\usepackage{xcolor}

\usepackage[pdftex]{graphicx}
\usepackage{dcolumn}
\usepackage{bm}

\usepackage[utf8]{inputenc}
\usepackage[T1]{fontenc}
\usepackage{mathptmx}
\usepackage{etoolbox}

\makeatletter
\def\@email#1#2{%
 \endgroup
 \patchcmd{\titleblock@produce}
  {\frontmatter@RRAPformat}
  {\frontmatter@RRAPformat{\produce@RRAP{*#1\href{mailto:#2}{#2}}}\frontmatter@RRAPformat}
  {}{}
}%
\makeatother

\usepackage[columnwise]{lineno}

\begin{document}

\preprint{AIP/123-QED}

\title{24-hour measurement of squeezed light \\using automated stable fiber system}

\author{Tomohiro Nakamura}
\altaffiliation{nakamura@alice.t.u-tokyo.ac.jp}
\author{Takefumi Nomura}
\affiliation{Department of Applied Physics, School of Engineering, The University of Tokyo, 7-3-1 Hongo, Bunkyo-ku, Tokyo 113-8656,
Japan.}
\author{Mamoru Endo}
\affiliation{Department of Applied Physics, School of Engineering, The University of Tokyo, 7-3-1 Hongo, Bunkyo-ku, Tokyo 113-8656,
Japan.}
\affiliation{Optical Quantum Computing Research Team, RIKEN Center for Quantum Computing, 2-1, Hirosawa, Wako, 351-0198, Saitama, Japan.}
\author{Atsushi Sakaguchi}
\affiliation{Optical Quantum Computing Research Team, RIKEN Center for Quantum Computing, 2-1, Hirosawa, Wako, 351-0198, Saitama, Japan.}
\author{He Ruofan}
\affiliation{Department of Applied Physics, School of Engineering, The University of Tokyo, 7-3-1 Hongo, Bunkyo-ku, Tokyo 113-8656,
Japan.}
\author{Takahiro Kashiwazaki}
\author{Takeshi Umeki}
\affiliation{NTT Device Technology Labs, NTT Corporation, 3-1, Morinosato Wakamiya, Atsugi, 243-0198, Kanagawa, Japan.}
\author{Kan Takase}
\author{Warit Asavanant}
\affiliation{Department of Applied Physics, School of Engineering, The University of Tokyo, 7-3-1 Hongo, Bunkyo-ku, Tokyo 113-8656,
Japan.}
\affiliation{Optical Quantum Computing Research Team, RIKEN Center for Quantum Computing, 2-1, Hirosawa, Wako, 351-0198, Saitama, Japan.}
\author{Jun-ichi Yoshikawa}
\affiliation{Optical Quantum Computing Research Team, RIKEN Center for Quantum Computing, 2-1, Hirosawa, Wako, 351-0198, Saitama, Japan.}
\author{Akira Furusawa}
\altaffiliation{Corresponding Author : akiraf@ap.t.u-tokyo.ac.jp}
\affiliation{Department of Applied Physics, School of Engineering, The University of Tokyo, 7-3-1 Hongo, Bunkyo-ku, Tokyo 113-8656,
Japan.}
\affiliation{Optical Quantum Computing Research Team, RIKEN Center for Quantum Computing, 2-1, Hirosawa, Wako, 351-0198, Saitama, Japan.}

\date{\today}

\begin{abstract}
In order to provide a cloud service of optical quantum computing, it is inevitable to stabilize the optical system for many hours. It is advantageous to construct a fiber-based system, which does not require spatial alignment. However,  fiber-based systems are instead subject to fiber-specific instabilities. For instance, there are phase drifts due to ambient temperature changes and external disturbances, and polarization fluctuations due to the finite polarization extinction ratio of fiber components. Here, we report the success of measuring squeezed light with a fiber system for 24 hours. To do this, we introduce stabilization mechanics to suppress fluctuations in the fiber system, and integrated controller to automatically align the entire system. The squeezed light at the wavelength of 1545.3~nm is measured every 2 minutes, where automated alignments are inserted every 30 minutes. The squeezing levels with the average of $-4.42$~dB are recorded with an extremely small standard deviation of 0.08~dB over 24 hours.
\end{abstract}

\maketitle

\section{Introduction} \label{sec1}
To accelerate research and development of quantum computation, it is desired to open prototypes and allow researchers to access them. Google and IONQ have developed a high-fidelity 11-qubit ion-trap quantum computing accessible via Google Cloud\cite{Egan2021}. IBM has released a superconducting 127-qubit quantum processor ``Eagle'' in 2021 on the IBM Cloud\cite{ibm_quantum}. Xanadu has conceived and built ``Borealis'', a free-space-based quantum processor that can perform Gaussian boson sampling\cite{Madsen2022}, and made it available on their Cloud platform. Since ``Borealis'' is a free-space system except for some parts of the system, pointing of the optical beams drifts, which is recovered by frequent manual alignments. If the system is constructed with a fiber system instead of a free-space system, the frequent alignments are removed. Thus the fiber system is advantageous when constructing a quantum computer operating long time.

In particular, we are aiming to build a fiber-based optical quantum computer which operates on continuous variables of light. Quantum computation based on continuous variables is gaining attentions. Recently, large-scale entanglements using time-domain multiplexing and frequency-domain multiplexing have been demonstrated\cite{Raussendorf2001, Menicucci2006, Menicucci2011, Pysher2011, Yokoyama2013, Yoshikawa2016, Yang2016, Zhang2017, Larsen2019_2d, Asavanant2019, Pfister2020}. In time-domain multiplexing, large entanglement is generated by sending squeezed light to interferometers with asymmetric lengths of arms. 

However, if the optical system is constructed with fiber, fiber-specific problems arise. Phase drifts and polarization fluctuations are caused by the ambient temperature changes\cite{tur1995, Slavik2015, Lv2019}. For phase drifts, an optical delay of 10~m in free-space systems causes typically a few wavelengths of phase drift, while that in the fiber system, if no measures are taken, induces several hundred wavelengths of phase drift. For polarization drifts, even if polarization-maintaining fibers are used, polarization extinction ratio (PER) of fiber components is finite, and thus concatenating these devices leads to polarization instability. Furthermore, power fluctuations are induced by the polarization fluctuations when combined with polarization-depend components. In addition, the coupling ratio of fiber beamsplitter is slightly dependent on the polarization and temperature. Since squeezed light is degraded by optical loss, all fiber components through which the squeezed light passes should have low optical loss.

Here, we solve the above issues by introducing stabilization devices, and successfully measure squeezed light stably for 24 hours. For phase drifts, we cover the entire system with a wind shield to passively suppress large phase drifts, and then introduce the fiber stretchers to actively suppress the phase drifts with feed-back controls. For polarization fluctuations, active polarization controllers are introduced together with polarizers. In order to generate the error signal for the polarization control, a polarization modulation method is employed. For power fluctuations, an power stabilization mechanism is introduced. For a fiber beamsplitter whose coupling ratio variation is critical, we control temperature of the beamsplitter and stabilize the coupling ratio without introducing significant optical loss. Automated alignment of the entire system is achieved by connecting the various stabilization mechanisms via a network and integrating their control program. Using the above system, we measure the squeezed light in 1545.3~nm generated by the optical parametric amplifier (OPA) with periodically poled lithium niobate (PPLN) waveguide\cite{Kashiwazaki2020, Kashiwazaki2023a} at 2-minute intervals with automated alignment every 30 minutes. Squeezing levels with the average of $-4.42$~dB and the standard deviation of 0.08~dB are successfully measured over 24 hours. In this demonstration, the optical loss of squeezed light is avoided by placing the controllers in the paths where the squeezed light does not pass, except for the coupling ratio stabilization. On the other hand, in the case of a more complex interferometer built for optical quantum computers, there are situations where the polarization and the phase of squeezed light need to be controlled with low optical loss. In that case, the method proposed in Ref.\onlinecite{Nakamura2023} can be available.

There are previous works which demonstrate quantum optics experiments with fibers. In Ref.~\onlinecite{Takanashi2020b}, with a fiber packaged OPA, squeezed light is measured in fiber-based systems. In Ref.~\onlinecite{Larsen2019a}, squeezed light from free-space optical parametric oscillator is injected to fiber interferometers to generate two-dimensional cluster states. However, in the above two demonstrations, the polarization of light and the coupling ratio of fiber beamsplitters are controlled manually or uncontrolled. To realize long-time stability, alignment needs to be automated. On the other hand, long-time squeezed light measurements have been reported for free-space systems without beam pointing alignment\cite{Shajilal2022}. This previous experiment is working well because the system size of squeezed light measurement is relatively small. For more complex systems realizing optical quantum computer, fiber systems are advantageous as explained above, and thus our techniques introduced in this work can be applicable to such complex systems. 

This paper is structured as follows. We explain squeezed light in Section \ref{sec2}. We discuss why various stabilizations are needed in the experimental setup in Section \ref{sec3}. We illustrate an overview of the experimental system in Section \ref{sec4}. We explain stabilization of optical power in Section \ref{sec5}, phase lock in Section \ref{sec6}, polarization optimization in Section \ref{sec7}, and the coupling ratio stabilization of the fiber beamsplitter in Section \ref{sec8}. We show the results of the 24-hour squeezed light measurement in Section \ref{sec9}. Section \ref{sec10} is the conclusion.

\section{Squeezed light and measurement} \label{sec2}
\begin{figure}
\includegraphics[width=\linewidth]{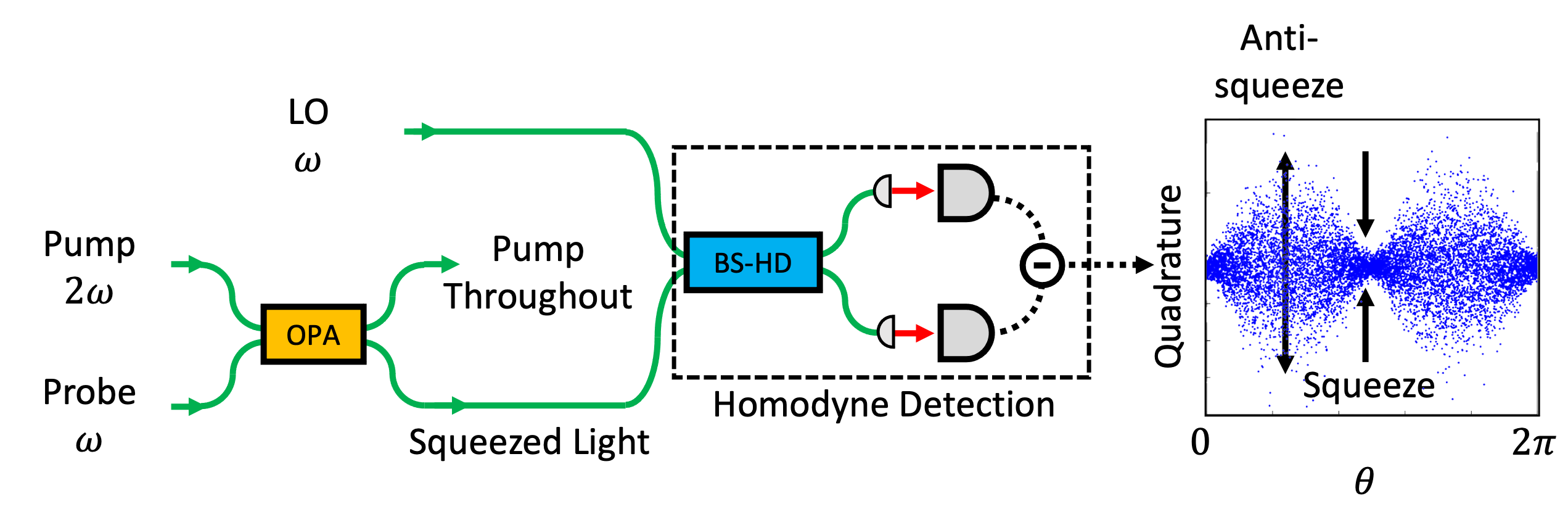}
\caption{Fundamental setup for squeezed light measurement. BS: beamsplitter; OPA: optical parametric amplifier; HD-BS: beamsplitter for homodyne detection; Green line: PM fiber; Dotted line: electric wire.}
\label{fig1}
\end{figure}

The quadrature phase amplitudes $\hat{x}$ and $\hat{p}$ of the single-mode quantized field of light have uncertainty $\Delta(\hat{x})\Delta(\hat{p}) \geq \hbar/2$, where $\hbar$ is the reduced Planck constant. The vacuum state is a minimum uncertainty state with $\Delta(\hat{x})=\Delta(\hat{p})=\sqrt{\hbar/2}$. A squeezed state has a quadrature phase amplitude where the fluctuation is smaller than that for a vacuum state\cite{Andersen2016}. For example, an $x$-squeezed state is squeezed as $\Delta(\hat{x}) < \sqrt{\hbar/2}$ and instead anti-squeezed as $\Delta(\hat{p}) > \sqrt{\hbar/2}$ to satisfy the uncertainty principle. Fault tolerant quantum computation can be achieved by using high-level squeezed states\cite{Menicucci2013, Fukui2018, Vahlbruch2016}. An arbitrary quadrature phase amplitude $\hat{x}\cos\theta+\hat{p}\sin\theta$ can be measured by setting the phase of the local oscillator (LO) light to $\theta$ in an optical homodyne detection. A vacuum state has a quadrature distribution insensitive to $\theta$, while a squeezed state has that sensitive to $\theta$. Figure~\ref{fig1}~(right) shows a typical quadrature distribution of a $x$-squeezed state, where the quadrature has the minimum variance at $\theta=0$, $\pi$, and the maximum variance at $\theta=\pi/2$, $3\pi/2$.

Figure~\ref{fig1} shows a simplified experimental setup for generating and measuring squeezed light. Here, the angular frequency of fundamental light is $\omega$. By injecting pump light with the angular frequency of $2\omega$ to a second-order nonlinear material, squeezed light is generated by parametric down conversion. In our demonstration here, squeezed light is generated by a PPLN waveguide OPA. In homodyne detection, squeezed light is combined with LO light at a 50:50 beamsplitter, and then is detected with two photodiodes in a homodyne detector, where the difference of the photocurrents shows the quadrature values. The squeezing and antisqueezing levels are obtained by comparing the variances with that of the vacuum state. The quadrature phase amplitude of the vacuum state is obtained by blocking the squeezed light and feeding only the LO light into the homodyne detector. Since the energy of a single photon corresponds to about 12,000~K in the telecommunication wavelength, thermal excitation is negligible at the room temperature and the vacuum state can be measured with this method.

It is necessary to lock the phase of the LO light depending on the measurement phase. For phase lock, coherent light of the fundamental wave is input to the OPA, and this light is referred to the probe light. Since the probe light undergoes phase-sensitive amplification which is in phase with the generated squeezed light, the relative phase between the probe light and the squeezed light can be locked. In homodyne detection, the interference signal between the probe light and LO light is locked, by which the relative phase between the squeezed light and the LO light is locked.

\section{Necessity for stabilization mechanics} \label{sec3}
Although we build the experimental setup for the squeezed light measurement, the stabilization mechanics which are implemented in this system is easily extended to the more complicated system.

In optical quantum computation, interference of squeezed light generates complicated entanglement structures. The polarizations must be matched and the relative phases must be locked to a specific direction at the interference points. If polarization and phase are not in the proper state, the quality of the entanglement is degraded. Even if the fiber components are made of polarization-maintaining fibers, connections between multiple fiber components shift the polarization state of the light from the polarization-maintaining axis. Birefringence variation caused by external disturbances fluctuates the polarization of light which is shifted from the polarization-maintaining axis. The polarization fluctuations also cause power fluctuations as the light passes through the polarization-dependent element. For example, fluctuations in LO power cause changes in the shot noise level in homodyne detections, which undermines the reliability of the experiment. Furthermore, changes in ambient temperature cause the fiber to expand and contract, which changes the optical path length, resulting in phase drift\cite{Elezov2018}. Besides, a fiber beamsplitter can be realized with a directional coupler, but its coupling ratio must be stable at an appropriate value.

In the squeezed light measurement experiments of this study shown in Fig.~\ref{fig1}, the issues of polarization fluctuation, power fluctuation, phase drift, and coupling ratio fluctuation of the beamsplitter appear as follows, and stabilization mechanics are installed. Regarding polarization, if the polarizations of the squeezed light and the LO light are misaligned, the squeezing level is degraded. Squeezed light is generated by OPA, but the polarization of the squeezed light is parallel to the crystal axis of PPLN waveguide. For the polarization alignment, the probe light is input to the path of the squeezed light. The polarization of the probe light is adjusted to match the polarization of the squeezed light, and then the polarization of the LO light is adjusted to match the polarization of LO light. Polarization controls are implemented with devices that change polarization by applying pressure to the fiber and causing birefringence. In the case of the LO light, power fluctuations change the shot noise level of the homodyne detection. In the case of the probe light, they change the amplitude of the error signals of the various locks, resulting in misalignment of the locking points. Therefore, it is necessary to stabilize the power of the LO and probe light, and fiber pigtailed variable attenuators are used. As for the phase, if the phase lock is not able to cancel the phase drift, an anti-squeeze component will be mixed when measuring the squeezing level, resulting in the reduction of squeezing level. In this experiments, fiber stretchers are inserted for the phase lock between the pump light and the probe light and for that between the probe light and the LO light. The coupling ratio of the fiber beamsplitter used in this study for homodyne detection has severe hysteresis when the knob is rotated, making it difficult to adjust it manually. Futhermore, the polarization of the input light also change the coupling ratio. Therefore, the coupling ratio is controlled by temperature.

\section{Outline of the experimental system} \label{sec4}
\begin{figure*}
\centering
\includegraphics[width=\linewidth]{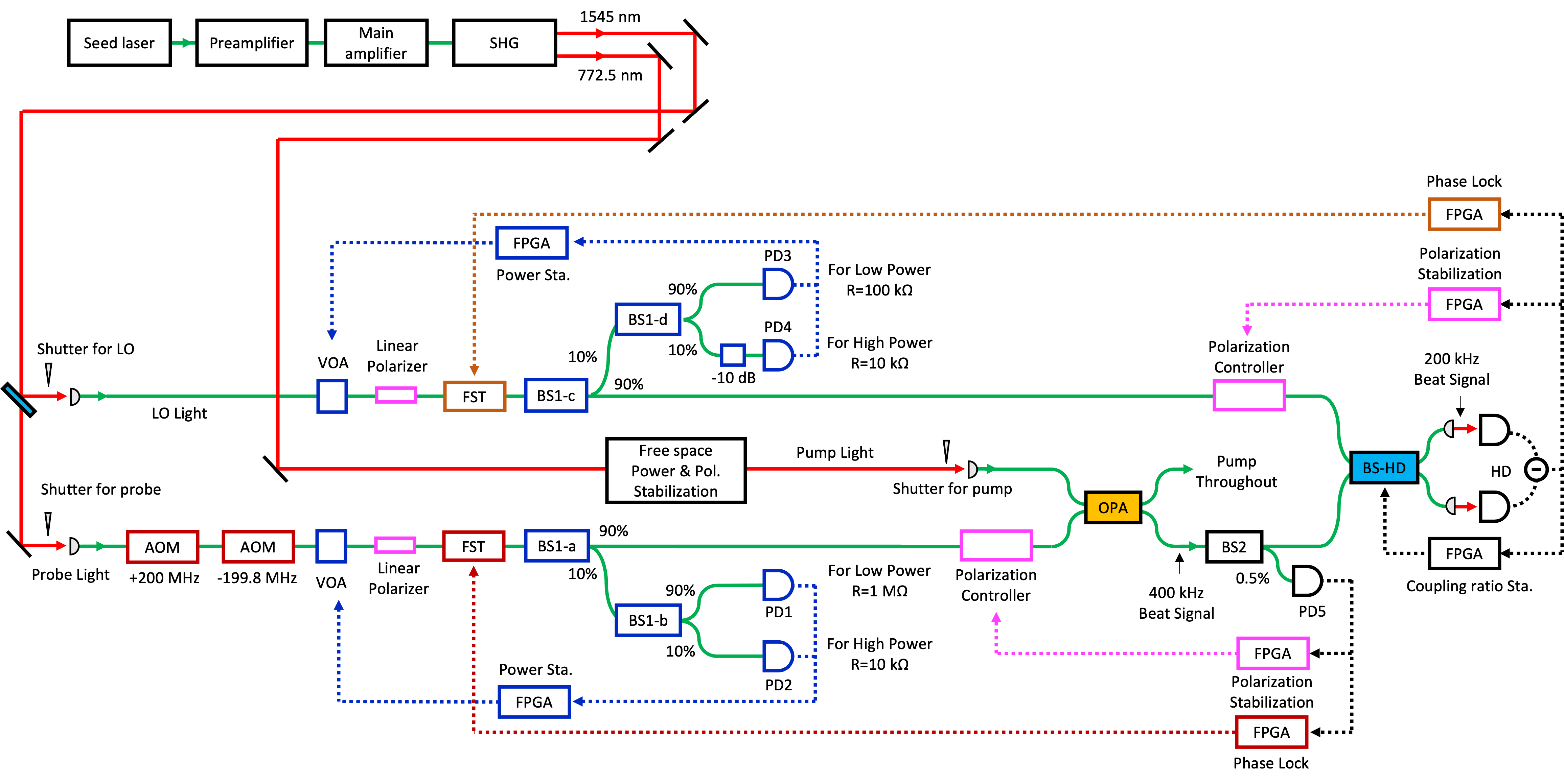}
\caption{The experimental system of this study. blue objects are for light power stabilization, pink objects are for polarization optimization, brown objects are for phase lock. AOM: acousto optic modulator; VOA: variable optical attenuator; FST: fiber stretcher; BS: beamsplitter; HD-BS: beamsplitter for homodyne detection; OPA: optical parametric amplifier; FPGA: field programmable gate array; PD: photo detector; HD: homodyne detector; Green line: PM fiber; Dotted line: electric wire.}
\label{fig2}
\end{figure*}

In this study, squeezed light is measured over 24 hours with the experimental system shown in Fig.~\ref{fig2}. In order to ensure stable measurements, light power stabilization, phase lock, polarization optimization, and coupling ratio lock of the beamsplitter are performed, which are described in the following sections. In this section, we describe the entire setup of the experimental system.

The laser source consists of multiple devices. The seed laser is Koheras ADJUSTIK X15 (NKT Photonics), whose wavelength and power are 1545.3~nm and 15~mW, respectively. The preamplifier, which reduces relative power noise by feedback, is Koheras BOOSTIK Linecard (NKT Photonics) whose output power of 80~mW is then attenuated to 20~mW. The main amplifier is Koheras BOOSTIK (NKT Photonics) whose output power is 8~W. The second harmonic generator (SHG) is Koheras HARMONIK (NKT Photonics), which converts some part of 1545.3~nm light to 772.7~nm light. After the SHG, the 1545.3 nm light and the 772.7~nm light are emitted to free space, which are then coupled to fibers. The 1545.3~nm light is used as the probe and LO light and the 772.7 nm light as the pump light. Before the fiber coupling, motor-driven optical shutters are inserted for all light paths to block the light, which is utilized in the sequence of control explained later.

Fiber components for the probe, the LO and the pump light are as follows. For the probe path, two acousto-optic modulators (AOM) SGTF200-1550-1P (Chongqing Smart science\&Technology Development) are inserted. Then, for the probe and the LO light paths, a fiber-based variable optical attenuator for power stabilization, a fiber stretcher for phase control, a polarization controller are connected. In the pump light path, the light power is stabilized by a motorized waveplate and polarizing beamsplitter in free space. The probe light and the pump light are input to the OPA, which is a PPLN waveguide and packaged with fiber input and output. The PPLN waveguide inside the OPA is phase-matched with type 0. Only the component parallel to the crystal axis of the input pump light contributes to the generation of squeezed light at the PPLN waveguide. This means that polarization fluctuations of the pump light do not cause polarization fluctuations of the squeezed light. Therefore, for the pump light, only the power stabilization is sufficient, which is performed by a free-space motorized waveplate and a polarizing beamsplitter. In the demonstration in Section \ref{sec9}, the power of the pump light is set to be 300~mW.

In the homodyne detection, the probe light interferes with the LO light at a fiber beamsplitter with a coupling ratio of 50:50, and are sent to the homodyne detector. Since the utilized photodiodes are for the free-space light, the light is output to free space by a collimator and focused on the photodiode by a concave mirror with the radius of curvature of 100~mm. For all-fiber implementation, the photodiodes can be fiber pigtailed in the future. The photodiodes of the homodyne detector are high quantum efficiency InGaAs-OD@1550nm (Laser Components) with a measured quantum efficiency of 96\%. The difference of photocurrents is converted to voltage signal by a transimpedance amplifier, which is composed of OPA847 (Texas Instruments) and 3~k$\Omega$ resistance. The signal from the homodyne detector is used for two purposes. The first is to generate a error signal for the phase lock, the polarization optimization, and the coupling ratio lock of the 50:50 beamsplitter. The second is to measure the squeezing level. The signal for the squeezing level measurement is high-pass filtered with the cutoff frequency of 15~MHz, and then is amplified by a factor of 15.

Because squeezed light is degraded by optical loss, optical components through which the squeezed light passes should have low optical loss. The fiber beamsplitters BS2 and BS-HD are 954P (Evanescent optics), which have loss of less than 0.1~dB. The fiber components are made of polarization maintaining fibers except for the polarization controller.

For the power stabilization, the phase lock, the polarization optimization and the coupling ratio stabilization of beam splitter, a number of field-programmable gate arrays (FPGAs), STEMlab 125-14 RedPitaya, are utilized, and they are controlled by a single computer via the socket communication. Multiple modulation and demodulation signals are required for phase lock and polarization optimization, which are synchronously generated by a direct digital synthesizer (AD9959, Analog Devices).

\section{Light power stabilization} \label{sec5}
In Fig.~\ref{fig2}, components which serve for power stabilization are colored in blue. To monitor the power fluctuation, some part of the light is picked up and fed back to the variable optical attenuator (NEW MMVOA-1-1550-P-8/125-SCSC-1-0.5, OZ Optics). To pick the light up, beamsplitters BS1-a, BS1-c (PMFC-1x2-1550-10/90-B-900-5-1-SC-P25, Opneti) with coupling ratio of 90:10 are used. To realize automated alignment, the light power needs to be stabilized at different orders of magnitude during alignment and measurement sequence. The target powers of the probe light are 1~$\mu$W for the OPA alignment, 100~$\mu$W for homodyne alignment, 100~nW for squeezed light measurement. The target powers of the LO light are 100~$\mu$W for alignment and 16~mW for measurement. For the low power, it is necessary to largely amplify weak light, and for the high power, it is necessary to reduce the amount of light to avoid the saturation of the PD. Therefore, the picked-up light is further divided by a 90:10 beamsplitter BS1-b, BS1-d, whose model number is the same as that of BS1-a and BS1-c. Then, the divided light goes to the two photo detectors (PDs), one for the high power and one for the low power. PD1, PD2, PD3 and PD4 contain a photodiode G8195-11 (Hamamatsu Photonics) and a transimpedance amplifier, where the transimpedance is 1~M$\Omega$, 10~k$\Omega$, 100~k$\Omega$ and 10~k$\Omega$, respectively. Before PD4, a 10~dB attenuator is inserted. Because the finite PER of the fiber components can cause the polarization change, the PER of BS1-a,b,c,d is ordered to be better than 25~dB. Without the power stabilization, the optical power drift was 15\% at most, but with the power stabilization on, the standard deviation of the power fluctuation was 0.1\%.

\section{Phase lock} \label{sec6}
In Fig.~\ref{fig2}, components that serve for phase locks are colored in blue. There are two phase lock points: the parametric amplification of the OPA (the phase between the probe light and the pump light) and the homodyne detection (the phase between the probe light and the LO light) . The probe light is frequency-shifted by the AOMs before the OPA. The frequency is up-shifted by 200.0~MHz at the first AOM, and then down-shifted by 199.8~MHz at the second AOM, resulting in a total frequency shift of 200~kHz. At the OPA, the probe light is parametrically amplified depending on the phase between the probe light and the pump light. The parametric amplification generates $-200$~kHz detuned light from the 200~kHz detuned probe light, thus resulting in the a 400~kHz beat signal. The probe light is picked up by a fiber beamsplitter BS2 with a coupling ratio of 99.5:0.5, and the 400~kHz beat signal for the phase lock of the OPA is acquired with a PD5 using an avalanche photo diode (KPDEA007-T, Kyoto Semiconductor) followed by an amplification of $M$=10. At the homodyne detector, since the LO light is not frequency-shifted, a beat signal of 200~kHz is observed, which is used for the phase lock of the homodyne detection. The beat signals obtained by the PD5 and the homodyne detector are demodulated, and the error signals are fed back to the fiber stretchers (FST). The fiber stretchers have the following structure. A cube-shaped piezoelectric actuator PC4FL (Thorlabs) is sandwiched by two 25~mm-diameter semicircle metals, and a polarization-maintaining fiber is wound by 10 times around it. The dynamic range of the fiber stretcher is about 3 wavelengths when the applied voltage range is set to 70~V. If the input polarization state deviates from the polarization-maintaining axis of the polarization-maintaining fiber, the polarization changes depending to the pressure to the fiber. Therefore, the fiber stretcher is placed immediately after the linear polarizer. The entire experimental system is covered by an airtight windshield made of aluminum frames, aluminum plates, and polycarbonate plates to suppress severe phase drifts due to temperature changes. The phase lock precision of the OPA was about 2$^\circ$ and that of the homodyne detection was about 0.3$^\circ$.

\section{Polarization optimization} \label{sec7}
In Fig.~\ref{fig2}, components that serve for the polarization optimizations are colored in pink. First, the polarizations of the probe light before the OPA and the LO light are stabilized by linear polarizers (ILP-1550-900-1-0.5-SA-P30, Opneti) with the PER$>$30 dB. Note that since squeezed light is sensitive to optical loss, the polarization stabilization should not be performed by inserting a linear polarizer in the path of the squeezed light after the OPA. The polarizations of the probe light and the LO light after the linear polarizers are slightly changed unintentionally by the followed FST and BS1, which is critical in the squeezed light measurement. Therefore, we actively optimize the polarization using the fiber polarization controller PCD-M02 (LUNA), which can control polarization with three degrees of freedom.

The polarization optimization by feedback control is not a trivial problem. In the case of the LO, for instance, feedback signal which goes to the polarization controller can be adjusted so that the amplitude of the interference signal is maximized. To investigate whether the current polarization state is optimized, the polarization state needs to be shifted slightly and we need to check the decrease of the amplitude. A previous research implements the polarization optimization by random walk\cite{Huang2009}. However, this random walk method is vulnerable to light power fluctuations, which cause amplitude fluctuations of the interference signal. Even if the polarization state is already in the optimal state, the control system may judge that the polarization is not optimal due to the power fluctuations, and moves from the current polarization state.

To solve this problem, we introduce a polarization modulation method. By applying a modulation to the polarization and obtaining the error signal from the demodulation, it becomes possible to lock the polarization at the zero crosspoint of the error signal. For the phase-sensitive amplification in OPA, the polarization modulation is applied to the probe light. For the interference at BS-HD for homodyne detection, it is applied to the LO light.

\subsection{Polarization modulation method} \label{sub7.1}
In this subsection, the polarization modulation method for the homodyne detection is mathematically described, and that for the OPA is described in Appendix~\ref{secA}. The polarization modulation is turned on only during polarization alignment and off during squeezed light measurement. The on/off sequence of polarization control is described in Appendix ~\ref{secB}.

\begin{figure}
\centering
\includegraphics[width=0.5\linewidth]{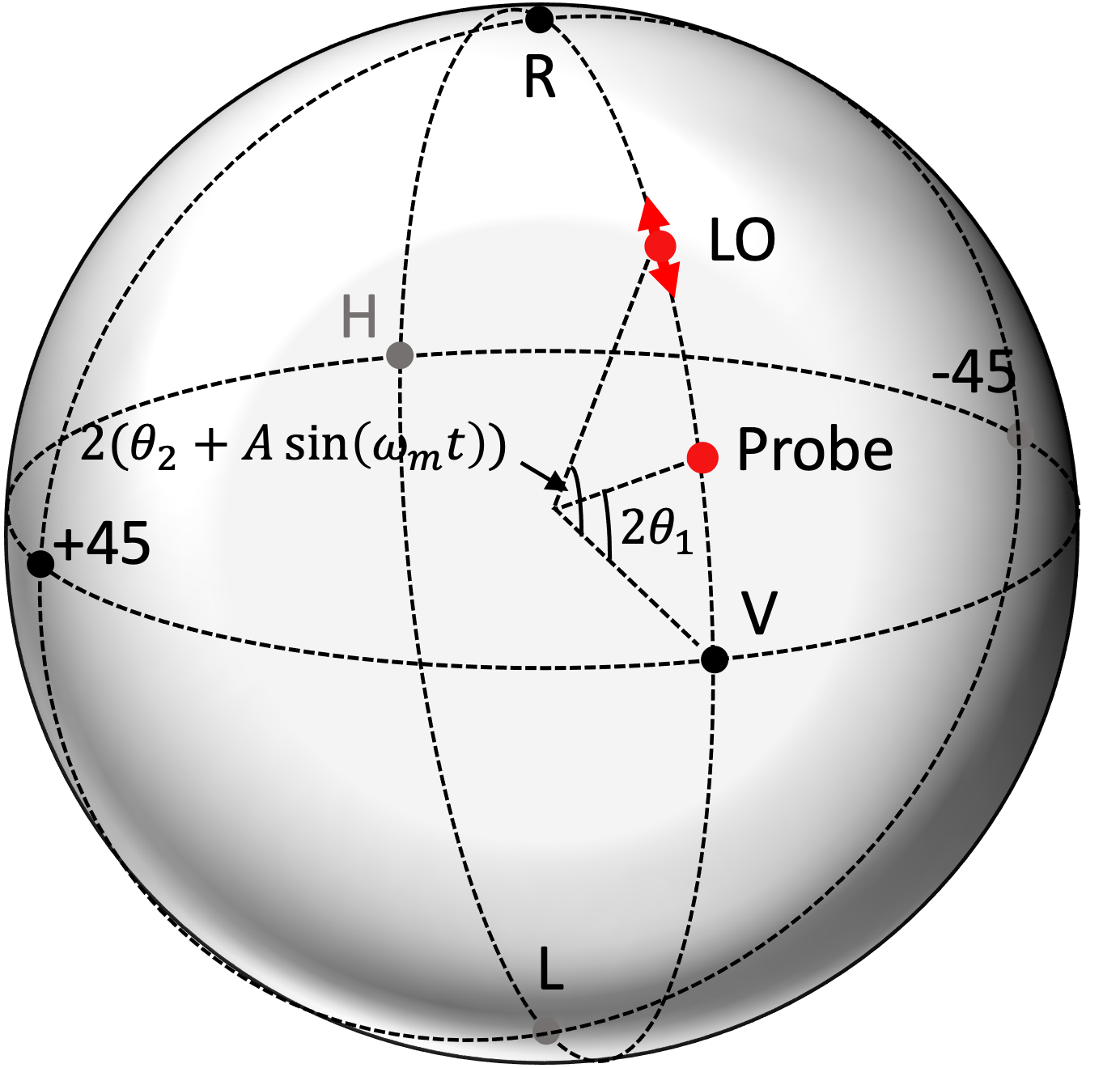}
\caption{The polarization of the probe light and the LO light on the Poincar\'{e} sphere in the homodyne detection. The polarization modulation is applied only to the LO light. On the Poincar\'{e} sphere, R, L,V, H, $+45$, $-45$ signs represent right-handed circular polarization, left-handed circular polarization, vertical linear polarization, horizontal linear polarization, linear polarization tilted by 45 degrees, and $-45$ represents linear polarization tilted by $-45$ degrees, respectively. The right-handed and left-handed circular polarization are defined from the point of view of the receiver.}
\label{fig3}
\end{figure}

We are aiming to match the polarizations of the probe light and the LO light at BS-HD by the polarization modulation. Although the polarization of the probe light can change during the experiment, a polarization controller with large loss should not be inserted in the path of the probe light through which the squeezed light is passing. Therefore, we put the polarization controller in the path of the LO light and align the polarization of the LO light to match that of the probe light. Note that we can instead insert a polarization controller with low optical loss to control the polarization of the squeezed light\cite{Nakamura2023}.

The polarization is represented by a Poincar\'{e} sphere, where the radius of the Poincar\'{e} sphere is normalized by 1. Here, for simplicity, the polarization of the probe light and the LO light lie on a great circle which passes through the polarization V and the polarization R on the Poincar\'{e} sphere, as shown in Fig.~\ref{fig3}. The polarization of the probe light is at the position rotated by $2\theta_1$ from the polarization V, and that of the LO light by $2\theta_2$. We assume the polarization controller can change the polarizations along this circle\cite{Shi2006}. General situations of the polarizations are discussed later. The polarization modulation is applied by the polarization controller in the path of the LO light. The polarization modulation is expressed by $2A\sin(\omega_m t)$, where $\omega_m$ is the angular frequency of the polarization modulation and $t$ is time. The angular frequency of the LO light is $\omega_0$ and the angular frequency of the probe light is shifted by $\Delta \omega$ from the LO light ($\Delta \omega$ corresponds to 200~kHz which is mentioned in Section \ref{sec6}). Let the light travel along the $z$ axis and the polarization V of the Poincar\'{e} sphere corresponds to the $x$ axis. Let $\mathbf{e}_x$ and $\mathbf{e}_y$ be unit vectors in the $x$ and $y$ axis, respectively. Assuming that the electric field amplitude is $E_1$ for the probe light and $E_2$ for the LO light. After they interfere by a polarization independent beamsplitter with a coupling ratio of 50:50, the complex electric fields of two output beams $\bm{E}_\text{HD}^{+}(t)$ and $\bm{E}_\text{HD}^{-}(t)$ are
\begin{align}
& \bm{E}_\text{HD}^{\pm}(t) \nonumber\\
&= \frac{1}{\sqrt{2}}E_1 (\cos\theta_1\mathbf{e}_x-i\sin\theta_1\mathbf{e}_y)\exp [-i(\omega_0+\Delta \omega) t+i\phi_1]\nonumber\\
& \quad \pm\frac{1}{\sqrt{2}}E_2\{\cos[\theta_2+A\sin(\omega_m t)]\mathbf{e}_x-i\sin[\theta_2+A\sin(\omega_m t)]\mathbf{e}_y\}\nonumber\\
& \quad\quad \exp(-i\omega_0 t+i\phi_2),\label{eqn1}
\end{align}
where $\phi_1$ and $\phi_2$ are the phase of the probe and LO light, respectively, which may drift during the experiments due to a external disturbances. The signals which are obtained by the following photodiodes are
\begin{align}
& |\bm{E}_\text{HD}^{\pm}(t)|^2 \nonumber\\
&= \frac{1}{2}E_1^2 + \frac{1}{2}E_2^2 \nonumber\\
& \quad \pm E_1E_2\cos[\theta_1-\theta_2-A\sin(\omega_m t)]\cos(-\Delta \omega t+\Delta\phi),\label{eqn2}
\end{align}
where $\Delta\phi = \phi_1-\phi_2$. With the homodyne detector, we measure the difference between the signals obtained from two photodiodes 
\begin{align}
& |\bm{E}_\text{HD}^{+}(t)|^2-|\bm{E}_\text{HD}^{-}(t)|^2 \nonumber\\
&=2E_1E_2\cos[\theta_1-\theta_2-A\sin(\omega_m t)]\cos(-\Delta \omega t+\Delta\phi),\label{eqn3}
\end{align}
where the constant term is removed. Here, we assume $A \ll 1$ and use the following approximation;
\begin{equation}
\cos[\theta-A\sin (\omega_m t)] = \cos\theta+A\sin(2\theta)\sin(\omega_m t). \label{eqn4}
\end{equation}
By applying the above approximation, Eq.\eqref{eqn3} becomes
\begin{align}
& |\bm{E}_\text{HD}^{+}(t)|^2-|\bm{E}_\text{HD}^{-}(t)|^2 \nonumber\\
&=2E_1E_2 [\cos(\theta_1-\theta_2)+A\sin(\theta_1-\theta_2)\sin(\omega_m t)]\nonumber\\
&\qquad \cos(-\Delta \omega t+\Delta\phi).\label{eqn5}
\end{align}
From the above equation, the beat signal with the frequency $\Delta \omega$ is modulated at the frequency $\omega_m$ with a coefficient $A\sin(\theta_1-\theta_2)$. Here, $\Delta \omega$ is chosen to be much higher than $\omega_m$ in order to remove the beat signal at the frequency of $\Delta \omega$ by a square-law detection. The square-law detection enables to neglect the phase drift $\Delta\phi$, in contrast to the demodulation at the frequency of $\Delta \omega$, which requires the synchronization of the demodulation signal depending on the phase drift of $\Delta\phi$. Then $\omega_m$ is demodulated, and finally the error signal proportional to $\sin(\theta_1-\theta_2)$ is obtained. This error signal is fed back to stabilize the polarization at $\theta_1 = \theta_2$. Note that Eq.~\eqref{eqn5} and Eq.~\eqref{eqn14} in Appendix \ref{secA} are expressed in the same formula
\begin{equation}
V_0(1+\beta \cos(\Omega t))\cos(\omega_c t+\Delta\phi). \label{eqn6}
\end{equation}
Here, $\beta$ indicates how much the polarization deviates from the optimal polarization state ($\beta=0$). By extracting $\beta$, both polarizations at the OPA and the homodyne detection can be optimized in the similar manner.

So far we considered optimization in one degree of freedom, but the surface on the Poincar\'{e} sphere has two degrees of freedom. It is widely known that rotations around three axes are able to realize arbitrary change of the polarization. For example, polarization controllers we used in this experiment have three piezo actuators in line, Piezo~1, Piezo~2, and Piezo~3. Those piezo actuators apply pressure to a single-mode fiber (not a polarization maintaining fiber) to change the output polarization. Piezo~2 is tilted by 45 degrees relative to the others, and this configuration enables an arbitrary polarization change\cite{Shi2006}. The polarization modulation must be applied by the same piezo actuator used for the feedback control, and thus polarization modulation is also switched in turn.

\subsection{Implementation} \label{sub7.2}
In the actual experiments, we used the following parameters. Since the polarization controllers used in this demonstration has a bandwidth of 2~kHz, the frequency of the polarization modulation is set to 300~Hz. This frequency is too low to demodulate by analog electric circuits. Therefore, we use an FPGA, which can easily implement many filters for low frequency signals. Furthermore, the parameters for the filters can be changed on site. Figure~\ref{fig5} shows the circuit configuration inside the FPGA to control the polarization of the LO light at the homodyne detection. Since the frequency of the probe light is shifted by 200~\text{kHz}, the interference between the probe light and the LO light generates 200~\text{kHz} beat signal. In Process~1, the 200~kHz signal is extracted by a band-pass filter to improve the signal-to-noise ratio of the beat signal. In Process~2, square-law detection is applied to extract the power modulation component. After the signal is squared, the 400~\text{kHz} component is removed by a low-pass filter with a cutoff frequency of 30~\text{kHz}, and a high-pass filter with a cutoff frequency of 10~Hz is followed to remove the DC component. In Process~3, the signal is demodulated at a polarization modulation frequency of 300~Hz to extract $\beta$ in Eq.~\eqref{eqn6}. The 300-Hz signal for the demodulation goes through a high-pass filter with the cutoff frequency of 10~Hz to remove the DC component, and then an all-pass filter to change the phase. Then the power modulation signal which is obtained by Process~2 is mixed with the 300-Hz demodulation signal. The DC component corresponds to $\beta$, while the second harmonic signal at 600~Hz is also generated. The 600-Hz signal is removed by a band-rejection filter and the DC component is extracted by a low-pass filter with a cutoff frequency of 10~Hz. The sampling frequencies of the 200~kHz band-pass filter and the 30~kHz low-pass filter are 125~MHz, and those of other filters are downsampled to 1/16. The filter types are infinite response filter with second order. 

By the above process, the error signal $\beta$, which reflects deviation from the optimum polarization state, is obtained. Then the error signal is fed back so that the error signal goes to zero. The 10-Hz cutoff frequency of the last low-pass filters is sufficient as a feedback bandwidth for polarization control, because the polarizations vary slowly in the order of several tens of minutes caused by the ambient temperature change. 

\begin{figure}
\centering
\includegraphics[width=0.6\linewidth]{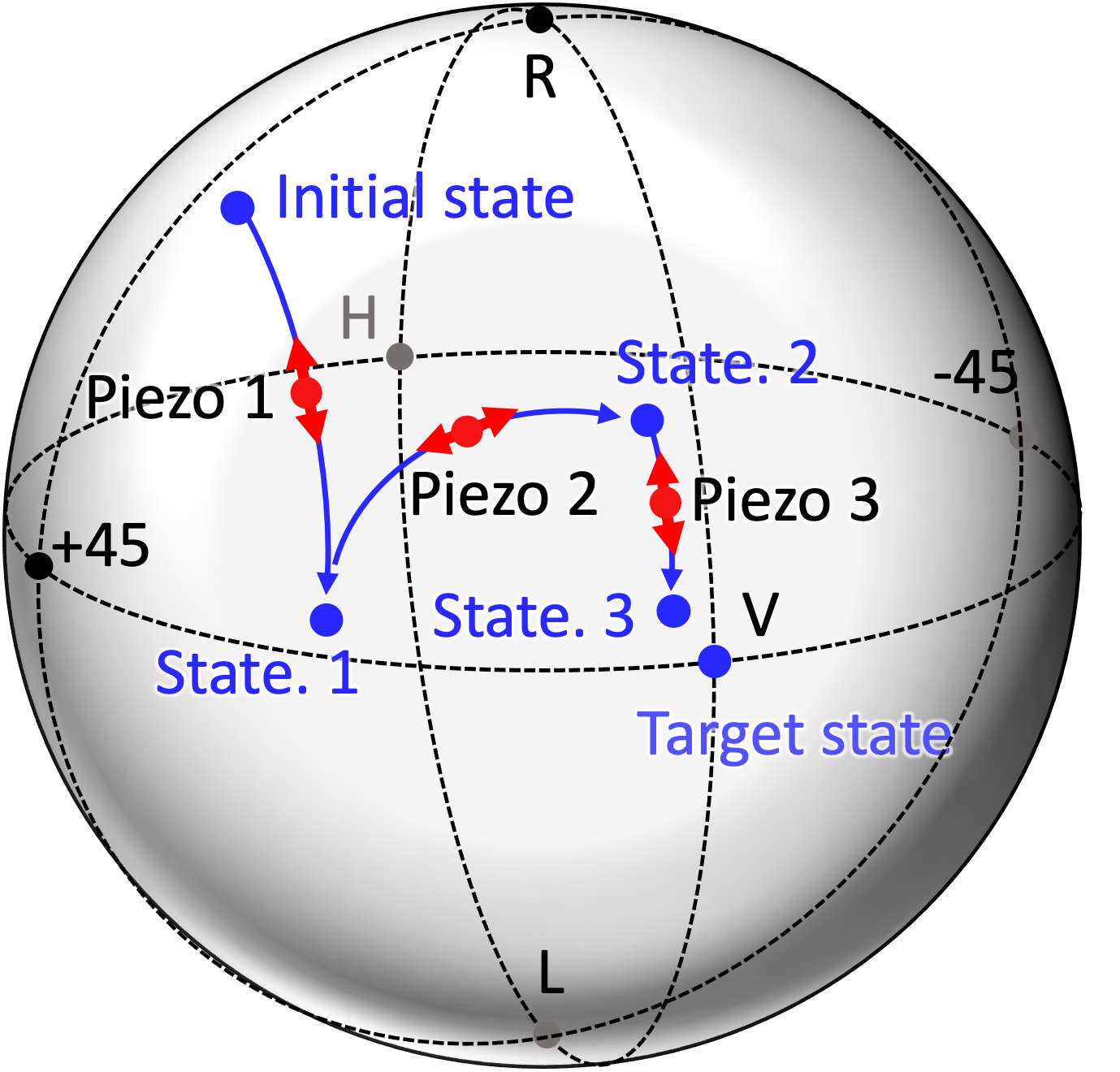}
\caption{Transition from the initial polarization state to the target polarization state using the polarization modulation method. Apply small polarization modulation to Piezo~1, Piezo~2, and Piezo~3 in turn and adjust the offset applied to them so that the error signal approaches $0$.}
\label{fig4}
\end{figure}

\begin{figure*}
\centering
\includegraphics[width=0.8\linewidth]{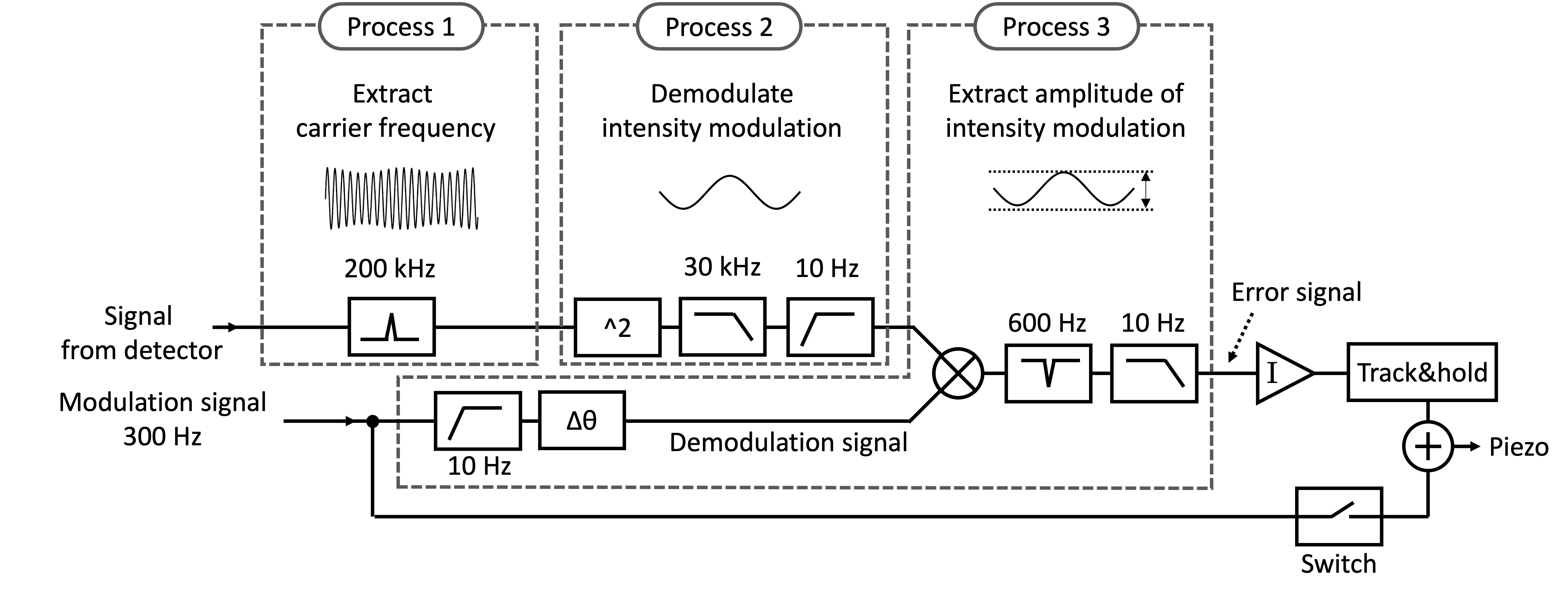}
\caption{The signal process inside the FPGA to generate the error signal from the power modulation signal. Symbol I means integral of the error signal. The switch cut the modulation signal. After the optimization, the voltage applied to the piezo actuator is held (track\&hold) and the modulation signal is cut.}
\label{fig5}
\end{figure*}

The piezo actuators inside the polarization controllers, to which the polarization modulation and the control signal are applied, accept the voltage range from 0~V to 140~V. If the digital input mode is selected, that voltage range corresponds to the integer range from 0 to 4095. The amplitude of the polarization modulation is set to 32.

The three piezo actuators are controlled in turn, as shown in Fig.~\ref{fig4}. First, Piezo~1 is vibrated to apply the polarization modulation, and the obtained error signal $\beta_1$ is integrated and then sent back to Piezo~1. When the error signal becomes close to zero, the feedback and the vibration of Piezo~1 is terminated. At this time, the polarization changes from the initial state to State~1. Next, similarly Piezo~2 is driven and the obtained error signal $\beta_2$ is processed, resulting in the polarization change from State~1 to State~2. Then, Piezo~3 and $\beta_3$ go in the same manner, resulting in the polarization change from State~2 to State~3. This process is cyclically repeated from Piezo~1 to Piezo~3 for several cycles to converge to the target state. Note that, since the polarization state may converge to the state opposite to the optimum depending on the initial state, the initial state should be manually set near the optimum.

\subsection{Comparison} \label{sub7.3}
\begin{figure*}
\centering
\includegraphics[width=\linewidth]{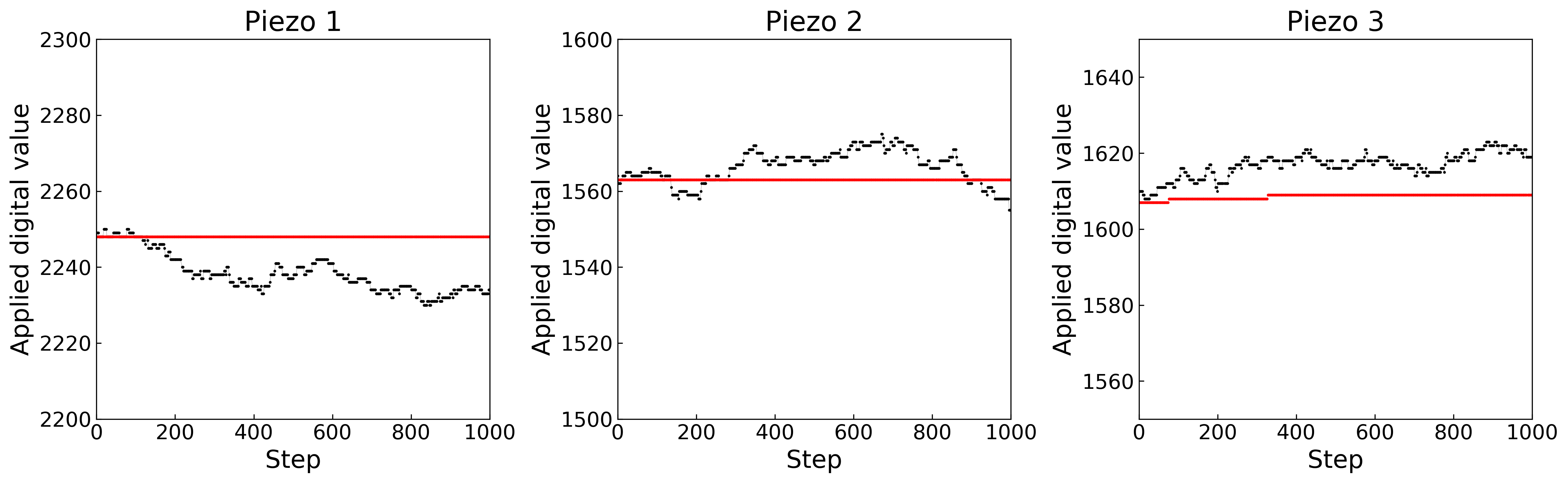}
\caption{Change in digital values sent to the three piezo actuators (Piezo~1, Piezo~2, and Piezo~3) when polarization optimization is performed with the random walk method (gray) and the polarization modulation method (red) for 1000 steps at 30 steps/s. In this figure, the traces for the polarization modulation method are the digital values which do not contain the modulation. The initial state is manually set to be the optimum polarization state. Since the polarization states fluctuate in the order of several tens of minutes, the digital values should not move in this time scale. For the random walk method, digital values are constantly fluctuating. On the other hand, for the polarization modulation method, they are stable.}
\label{fig6}
\end{figure*}

We made an experiment to compare the polarization modulation method with the random walk method. The initial state is set to the optimum polarization state by hand, and the polarization optimization is performed with 1000 feedback steps at a rate of 30 steps/s. Figure~\ref{fig6} shows the digital values sent to the three piezo actuators (Piezo~1, Piezo~2, and Piezo~3) for the random walk method (red) and the polarization modulation method (blue). Since the polarization state changes in the order of several tens of minutes, the polarization is expected to be stable in this measurement time scale (33 s). Even though the initial state is optimal, there are undesirable changes of the digital values for the random walk method. This is because, when the polarization is close to the optimum, slight power fluctuations degrade the stability of the random walk method, as already mentioned. On the other hand, the additional polarization drifts are not induced by the polarization modulation method. 

\section{Beamsplitter coupling ratio lock} \label{sec8}
Before explaining the beamsplitter coupling ratio lock, we refer to the structures of the beamsplitter. The beamsplitters are directional couplers, and light is exchanged between two fibers by evanescent coupling. Fiber beamsplitters are basically classified into two types: fused biconical tapered couplers\cite{Kawasaki1981, Pal2003} and side polished couplers\cite{Parriaux1981, Michel1982}. For the fused biconical tapered coupler, two fibers are twisted together and stretched on heating. In this method, the deformation of the cores results in the high optical loss. On the other hand, for the side-polished coupler, fibers are embedded in quartz substrates. Next, some part of the cladding is polished until it is removed to permit the evanescent coupling. Then the two pieces are contacted. In this method, the cores are not deformed, resulting in lower optical loss. To reduce the optical loss of squeezed light, we use side-polished fiber beamsplitters for the quantum light paths (BS2 and BS-HD in this demonstration).

In addition to the low-loss demand, the coupling ratio of the beamsplitter at the homodyne detection (BS-HD) must be exactly 50:50 for the following reason. In homodyne detection, the LO light interferes with the squeezed light and are divided at BS-HD. Then the difference of two photocurrents is electrically amplified, and the residual fluctuation represents the quadrature phase amplitude. Because the power of the LO light is very high, unbalance of the two photocurrents caused by a slight deviation of the coupling ratio from 50:50 easily results in saturation of the homodyne detector. In addition, since the DC output of the homodyne detector is used for generating various error signals as shown in Fig.~\ref{fig2}, we do not want to make the homodyne detector AC coupled to prevent saturation.

The coupling ratio is required to be 50:50 with an accuracy of less than 0.1\% from the following experimental parameters. The LO power is 16~mW, the probe power is 1~$\mu$W. The difference of the two photocurrents is converted to voltage by a transimpedance amplifer with the feedback resistance of 3.03~k$\Omega$, and then further amplified by 15 times. The output voltage of the detector should be kept in the range from $-1$~V to 1~V to avoid saturation of the op-amp.

\begin{figure}
\centering
\includegraphics[width=0.8\linewidth]{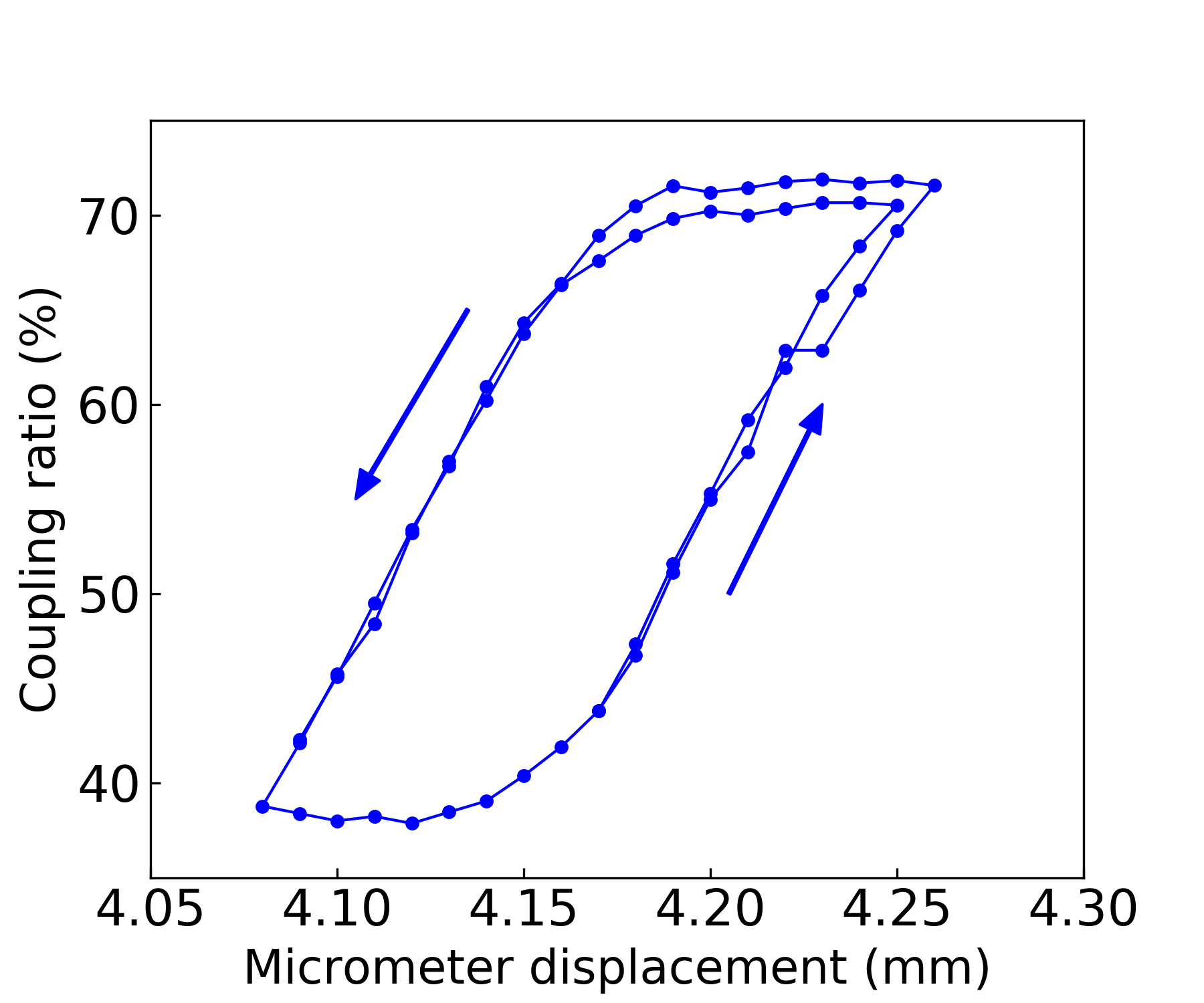}
\caption{Hysteresis of the coupling ratio depending on the micrometer displacement.}
\label{fig7}
\end{figure}

We use a fiber beamsplitter 905P (Evanescent Optics), whose coupling ratio can be changed by a micrometer. However, some obstacles are found for the stabilization of the coupling ratio. Firstly, the response of the coupling ratio has a hysteresis and a backlash for the micrometer displacement as shown in Fig.~\ref{fig7}. Secondly, it takes about 6 minutes for the coupling ratio to be stable after the user takes his or her hand off the knob. Thirdly, for the beamsplitter used in this study, the coupling ratio has dependency on the temperature at a rate of 0.95\%/${}^{\circ}$C around the coupling ratio of 50:50. Fourthly, the coupling ratio changes by about 2.9\% depending on the polarization state of the input light. Therefore, it is hard to maintain the coupling ratio with the accuracy of 0.1\%. We note that the above properties varies on the individuals.

\begin{figure}
\centering
\includegraphics[width=\linewidth]{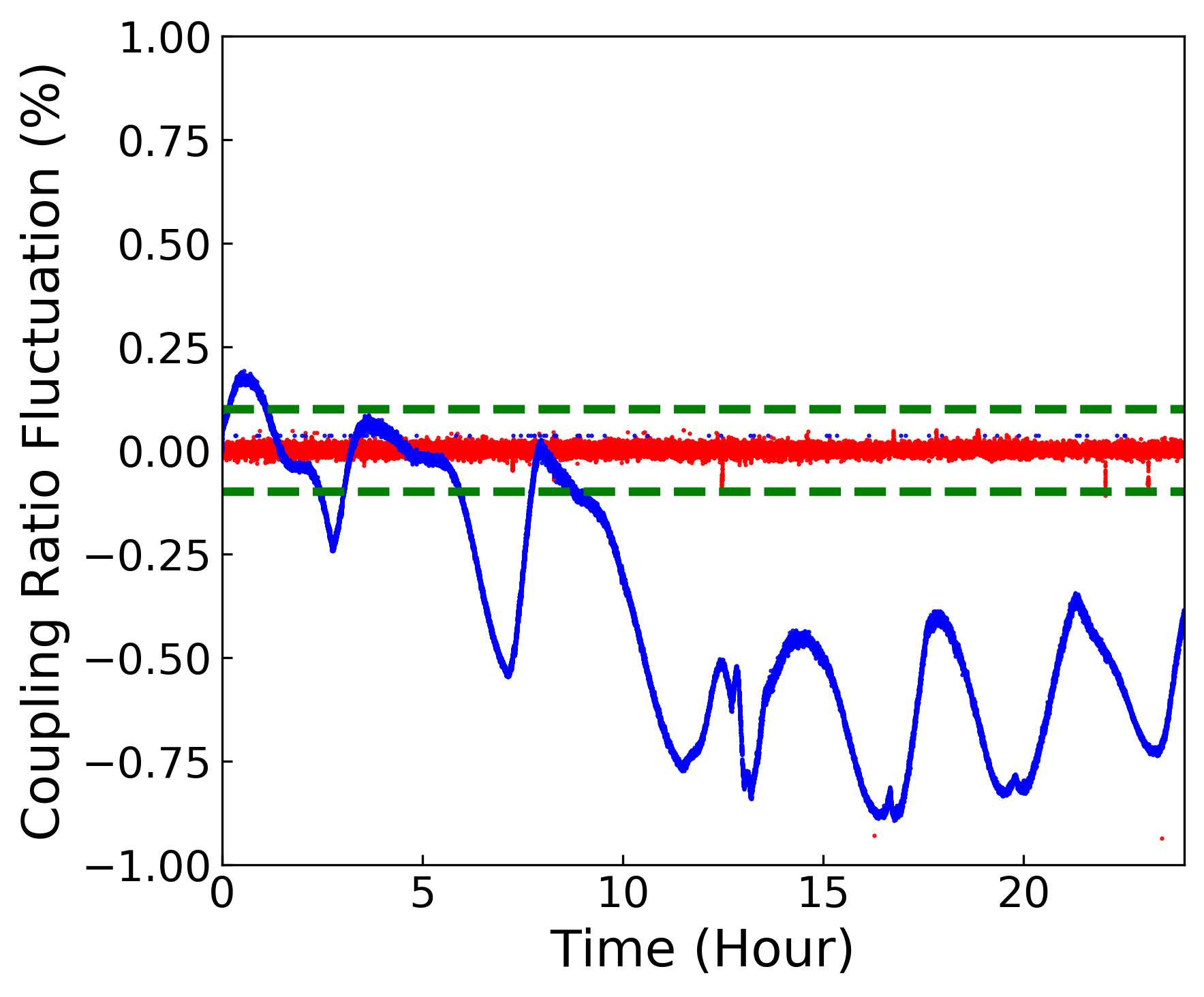}
\caption{Change of the coupling ratio of the fiber beamsplitter measured over 24 hours. Red: results when the homodyne signal is fed back to the Peltier element. Blue: results when the temperature measured by the thermistor is fed back to the Peltier element. Green: lines at the level of $\pm$0.1\%, where the homodyne detector does not saturate in this study. The standard deviation of the coupling ratio is 0.01\% for the red trace.}
\label{fig8}
\end{figure}

To stabilize the coupling ratio of the beamsplitter, we use the temperature as a degree of freedom for control. We attach a Peltier element to the bottom of the beam splitter and a thermistor to the side of the beam splitter, and place them on a heat sink. As a first test, we controlled the temperature measured by the thermistor to be kept constant. After 24 hours, the coupling ratio varied by about 1\% as shown in Fig.~\ref{fig8} (blue line), which was worse than the acceptable coupling ratio deviation of 0.1\%. Therefore, we employ a method to stabilize the coupling ratio of the beam splitter by monitoring the drift of the average of the homodyne detection signal and by feeding it back to the Peltier element. As shown in Fig.~\ref{fig8} (red line), the coupling ratio fluctuation is locked with a standard deviation of 0.01\% for 24 hours, which is sufficiently smaller than the target fluctuation of 0.1\%. Although this method use temperature dependence of the fiber beam splitter, some individuals of fiber beam splitter have little temperature dependence. Therefore, it is necessary to select individuals which has a sufficient temperature dependence or to develop a fiber beam splitter with a structure that has enough temperature dependence.

\section{24-hour squeezed light measurement} \label{sec9}
Squeezed light is measured for 24 hours in the fiber system using various stabilization mechanisms explained above. As described in Appendix~\ref{secB}, the alignments and the squeezed light measurements are repeated alternately in this experiment, with various controllers turned on and off in sequence. The spectra of the squeezing and antisqueezing levels in the range from 10~MHz to 200~MHz are acquired with resolution band width of 1~MHz and video band width of 10~Hz by a spectrum analyzer PXA Signal Analyzer N9030B (Keysight). 

\begin{figure*}
\centering
\includegraphics[width=\linewidth]{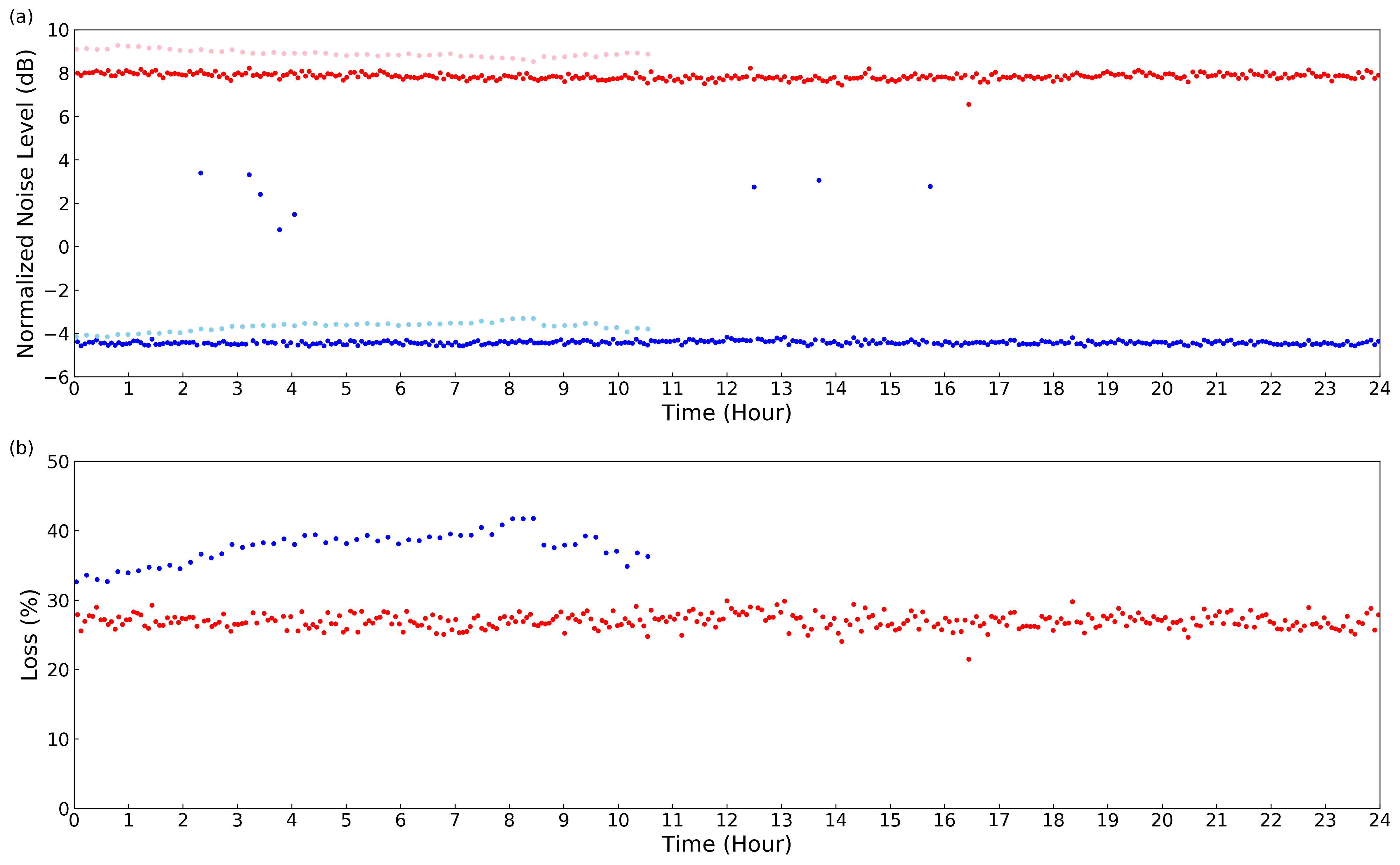}
\caption{Results of 24-hour squeezed light measurement. (a) Time variation of the squeezing and antisqueezing levels. Blue: squeezing level. Red: antisqueezing level. Sky blue: squeezing level without a part of control (reference). Pink: antisqueezing level without a part of control (reference). (b) Time variation of the effective loss of the entire experimental system. Red: loss with full control. Blue: loss without a part of control (reference).}
\label{fig9}
\end{figure*}

Figure~\ref{fig9}(a) shows time variations of squeezing level and antisqueezing level at 40~MHz as blue and red points, respectively. As a reference, an example of the squeezing level and antisqueezing level are shown as sky blue and pink points, respectively, for the case where the polarization stabilizations and coupling ratio lock of the beamsplitter are deactivated. Note that the phase locks and the power stabilizations are active for obtaining the sky blue and pink points. The blue and red points are obtained every 2 minutes, and the sky blue and pink points every 11 minutes. The blue points show an average squeezing level of $-4.42$~dB with an extremely small standard deviation $\pm$0.08~dB over 24 hours. The red points show an average antisqueezing level of 7.85~dB with a standard deviation 0.13~dB. In contrast, for the sky blue points, the squeezing level is not kept as the time passes. These results ensure that stabilization controls work properly to keep the system stable. Note that squeezing and antisqueezing levels have some outliers because they are not measured properly for a following reason. When the phase drift exceeds the dynamic range of the fiber stretcher, the voltage applied to the fiber stretcher is reset and the phase is relocked. If squeezed light measurements coincide with the reset process, inappropriate squeezing or antisqueezing levels are obtained. This problem can be prevented by programming of the control system, but we did not implement it in this measurement. When the average and the standard deviation of the squeezing and antisqueezing levels are calculated, these outliers are omitted.

We can estimate the effective loss of the system from the squeezing level and antisqueezing level. In addition to the insertion losses of optical components, the polarization mismatches become effective losses for the squeezed light, which reduces the squeezing level. If the change of the effective loss of the system is small, we can verify that the polarization matching is stable during the measurement. Figure~\ref{fig9}(b) shows the effective loss of the entire system estimated from the blue and red points in Fig.~\ref{fig9}(a). From the squeezing level $sq$ and antisqueezing level $asq$ expressed in dB, we can evaluate the effective loss $L$ of the whole system from the following relations,
\begin{align}
sq &= 10\log_{10}(L+(1-L)e^{-2r}), \label{eqn7}\\
asq &= 10\log_{10}(L+(1-L)e^{2r}), \label{eqn8}
\end{align}
where $r\ge 0$ is the squeezing parameter. In the ideal case where a loss is absent, the squeezing and antisqueezing levels become symmetric ($sq=-asq$). In actual cases where a loss exists, a vacuum fluctuation is mixed to the squeezed or antisuqeezed quadrature, from which the above equations are derived. From the above equations, $L$ is expressed as,
\begin{equation}
L=\frac{1-10^{\frac{sq}{10}+\frac{asq}{10}}}{2-10^{\frac{sq}{10}}-10^{\frac{asq}{10}}}. \label{eqn9}
\end{equation}
The time variation of effective loss was calculated to be $27.0\pm1.0$\% as shown in Fig.~\ref{fig9}(b). The fact that the loss fluctuation was only $\pm1.0$\% over 24 hours ensures that the control system is operating stably for a long period. The details of the 27\% loss are as follows. The loss of the OPA is from 8\% to 12\%. Two connectors are used, where the loss of each connector is from 3\% to 5\%. The loss of BS2 is about 3\%. The loss of BS-HD is about 0.5\%. The loss of the homodyne detection device is about 4\%. As an option, the connectors can be fused, by which the loss is reduced to 0.5\% per one connection point. In this experiment, fusion splicing is not done for the reusability of the OPA.

\begin{figure*}
\centering
\includegraphics[width=\linewidth]{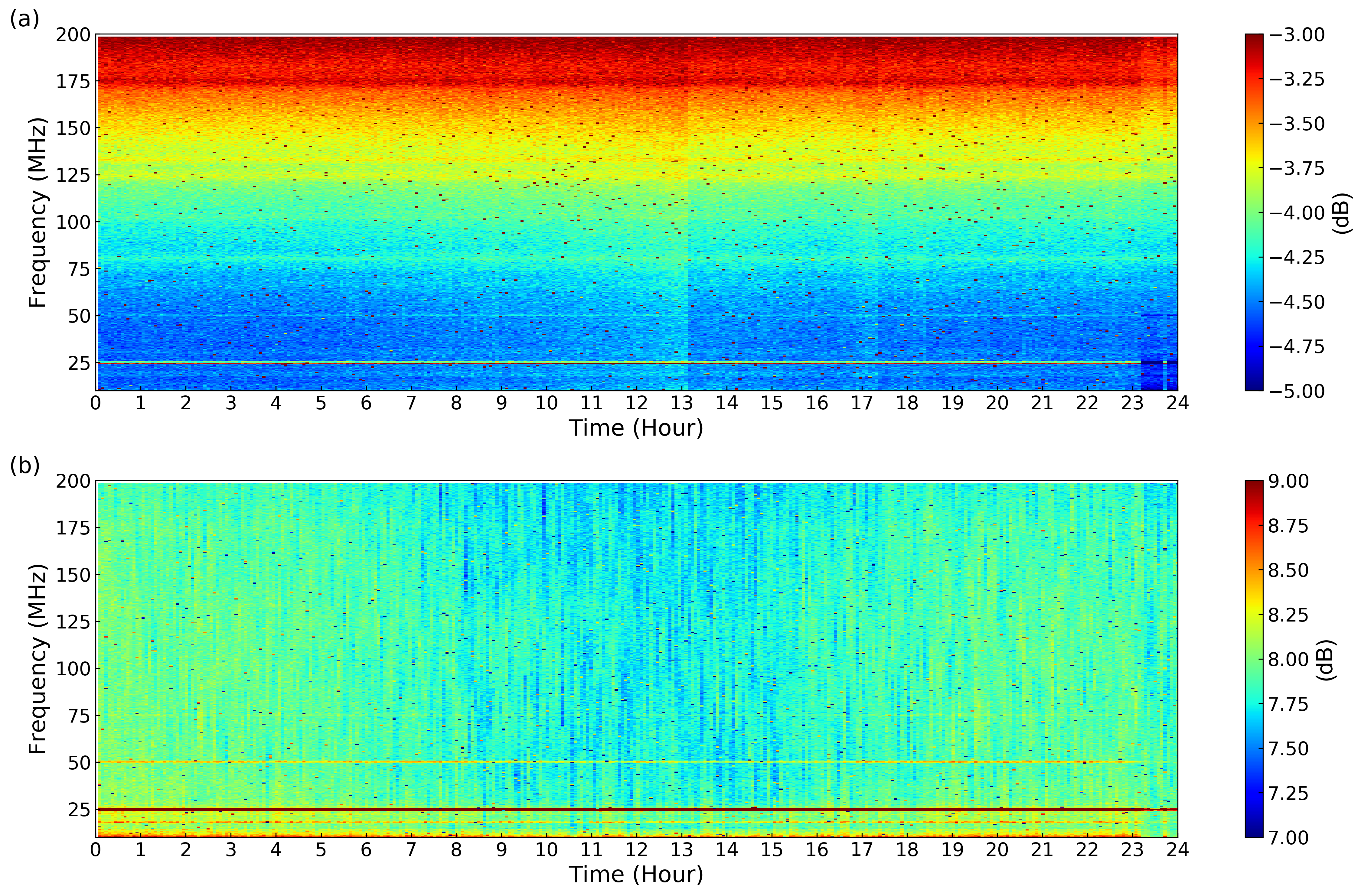}
\caption{Spectrogram from 10~MHz to 200~MHz for 24-hour squeezed measurements. (a) Squeezing level. (b) Antisqueezing level.}
\label{fig10}
\end{figure*}

Figure~\ref{fig10} shows a spectrogram of the squeezing and antisqueezing levels. The dots in Fig.~\ref{fig10} are due to the same reason of the phase re-locking as in Fig.~\ref{fig9}. The line at 25~MHz is due to electric noise.

\section{Conclusion} \label{sec10}
In this study, squeezed light with a squeezing level of $-4.42$~dB has been successfully generated and measured in a fiber-based experimental system with a very small standard deviation of $\pm0.08$~dB for 24 hours. Squeezed light is the most fundamental quantum state of light in cluster quantum computations\cite{Raussendorf2001, Menicucci2006, Menicucci2011, Pysher2011, Yokoyama2013, Yoshikawa2016, Yang2016, Zhang2017, Larsen2019_2d, Asavanant2019, Pfister2020} and loop quantum computations\cite{Takeda2017, takeda2019}. Based on the technology demonstrated in this paper, it will be possible to construct a system that is automatically maintained stable for long periods of time, enabling reliable operations of optical quantum computers offered as a cloud service.

\section{ACKNOWLEDGEMENTS} \label{sec11}
This work was partly supported by Japan Science and Technology Agency (Moonshot R\&D) Grants No.~JPMJMS2064 and No.~JPMJPR2254, Japan Society for the Promotion of Science KAKENHI Grants No.~18H05207 and Grants No.~20K15187, the UTokyo Foundation and donations from Nichia Corporation of Japan. T.N. acknowledges financial support from Forefront Physics and Mathematics Program to Drive Transformation (FoPM).  M.E. acknowledges support from Research Foundation for Opto-Science and Technology.

\appendix

\section{Polarization modulation at the OPA} \label{secA}

\begin{figure}
\centering
\includegraphics[width=0.5\linewidth]{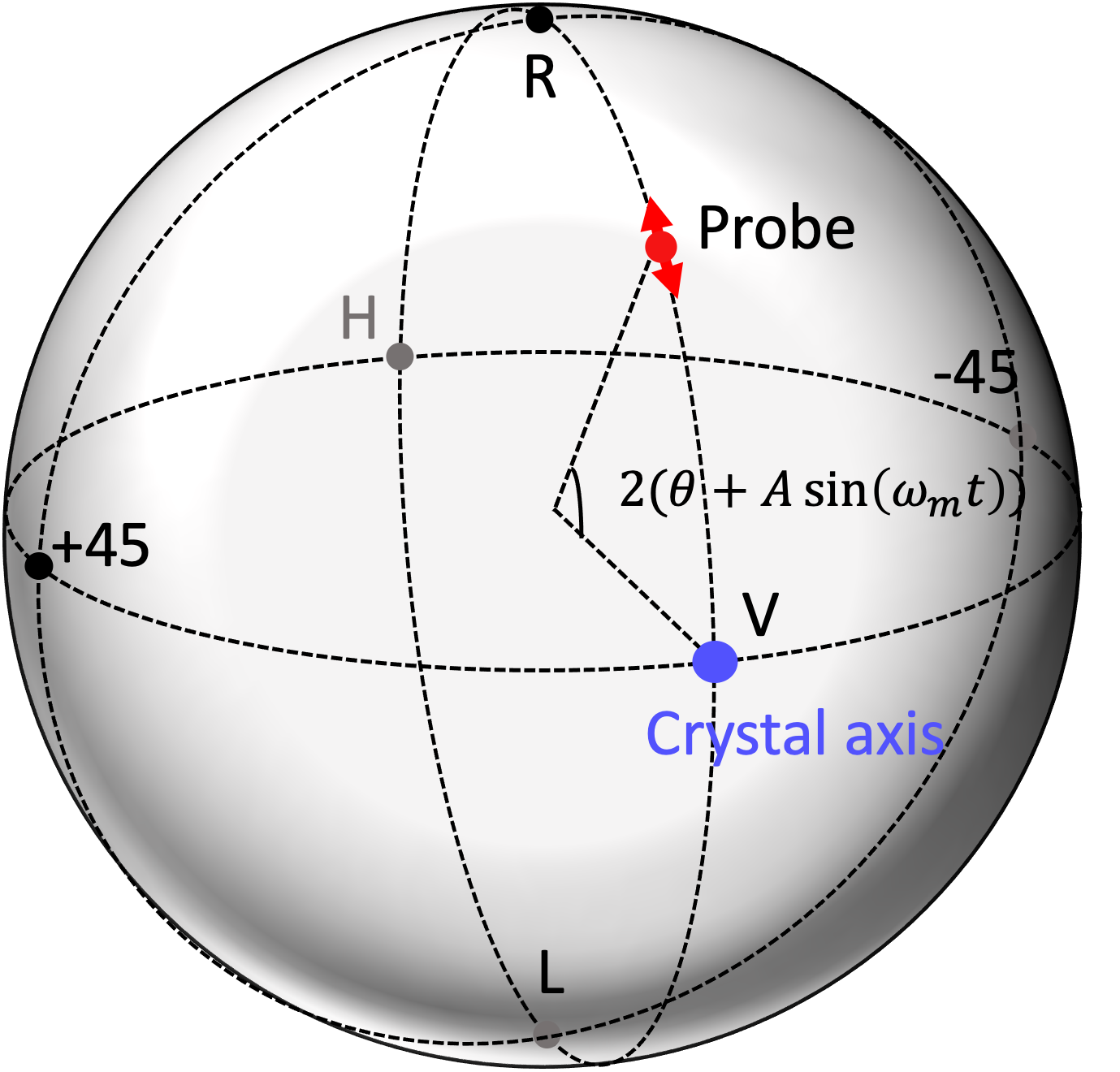}
\caption{Polarization on the Poincar\'{e} sphere of the probe light with polarization modulation before the OPA.}
\label{fig11}
\end{figure}

Here, we mathematically describe the polarization modulation method at the OPA. We aim to match the polarization of the probe light to the crystal axis where parametric down conversion happens. The crystal axis is set to the $x$ axis, which corresponds to the polarization V in the Poincar\'{e} sphere. In the similar manner to Section \ref{sec7} for simplicity, let us consider the situation where the polarization of the input probe light is shifted from the polarization V to the polarization R by $2\theta$ as shown in Fig.~\ref{fig11}. In this case, we aim to control $\theta$ to be $0$. The polarization of the probe light input to the OPA is slightly modulated on the Poincar\'{e} sphere by $2A\sin(\omega_m t)$, where $A$ is the amplitude of the polarization modulation ($A \ll 1$), $\omega_m$ is the angular frequency of polarization modulation, $t$ is time. The optical angular frequency of the degenerate parametric down conversion is denoted by $\omega_0$, the frequency shift of the probe light from the degenerate frequency $\Delta \omega$, the electric field amplitude $E_0$, and the optical phase of probe light $\phi$. Then the complex electric field of the probe light input to OPA $\bm{E}_\text{beforeOPA}$ is
\begin{align}
& \bm{E}_\text{beforeOPA} \nonumber\\
&=E_0\{\cos[\theta+A\sin(\omega_m t)]\mathbf{e}_x-i\sin[\theta+A\sin(\omega_m t)]\mathbf{e}_y\} \nonumber\\
& \quad\quad \exp[-i(\omega_0+\Delta \omega)t+i\phi_\text{probe}].\label{eqn10}
\end{align}
Only the polarization element of the probe light parallel to the crystal axis of a nonlinear crystal undergoes phase-sensitive amplification, while the perpendicular element remains unchanged. If the degrees of the phase-sensitive amplification is $\exp(r)$ and the phase of the pump light is $\phi_\text{pump}$, the electric field $\bm{E}_\text{afterOPA}$ of the probe light output from the OPA is\cite{Yariv1996}
\begin{align}
& \bm{E}_\text{afterOPA} \nonumber\\
&= E_0 \cos[\theta+A\sin (\omega_m t)] \nonumber\\
& \quad \times\{\cosh(r)\exp[-i(\omega_0+\Delta\omega) t+i\phi_\text{probe}]\nonumber\\
& \quad +i\sinh(r)\exp[-i(\omega_0-\Delta \omega) t-i\phi_\text{probe}+i\phi_\text{pump}]\}\mathbf{e}_x \nonumber\\
& \quad -iE_0 \sin[\theta+A\sin(\omega_m t)] \exp[-i(\omega_0+\Delta \omega)t+i\phi_\text{probe}] \mathbf{e}_y.\label{eqn11}
\end{align}
The intensity of light $I$ is proportional to the square of the electric field,
\begin{align}
I&\propto |\bm{E}_\text{afterOPA}|^2 \nonumber\\
&= E_0^2 \Bigl\{ \cos^2[\theta+A\sin(\omega_m t)]\nonumber\\
& \quad\quad  \times \Bigl[\cosh(2r)
+\sinh(2r)\sin(2\Delta \omega t-2\phi_\text{probe}+\phi_\text{pump})\Bigr]\nonumber\\
& \quad\quad +\sin^2[\theta+A\sin(\omega_m t)]\Bigr\}.\label{eqn12}
\end{align}
From $A \ll 1$, the following approximation is obtained.
\begin{equation}
\cos^2[\theta+A\sin (\omega_m t)] = \cos^2\theta-A\sin(2\theta)\sin(\omega_m t). \label{eqn13}
\end{equation}
By using the above expression, the element oscillating at angular frequency $2\Delta\omega$ in Eq.~\eqref{eqn12}, which is extracted by BPF, is
\begin{align}
&E_0^2[\cos^2\theta-A\sin(2\theta)\sin(\omega_m t)]\sinh(2r) \nonumber\\
& \quad \sin(2\Delta \omega t-2\phi_\text{probe}+\phi_\text{pump}). \label{eqn14}
\end{align}
The signal with the angular frequency of $2\Delta\omega$ is intensity modulated at an angular frequency of $\omega_m$. Intensity modulation is extracted by the square-law detection, where we can neglect the phase drift $-2\phi_\text{probe}+\phi_\text{pump}$, as explained in Section \ref{sec7}. The intensity modulation element is proportional to $\sin(2\theta)$, which is exacted by a demodulation at the frequency of $\omega_m$. By using $\sin(2\theta)$, we can control the polarization to $\theta=0$ or $\theta=\pi/2$, which depends on the polarity of the feedback.

The angular frequency of the carrier signal is $2\Delta \omega$ for the OPA as shown in Eq.\eqref{eqn14}, while $\Delta \omega$ for the homodyne detection as shown in Eq.\eqref{eqn5}. Thus, for the cascade filter implemented by FPGA, the center frequency of the BPF in Process 1 is 400~kHz for the OPA, while 200~kHz for the homodyne detection.

\section{Sequence for alignment and measurement} \label{secB}
\begin{figure*}
\centering
\includegraphics[width=0.8\linewidth]{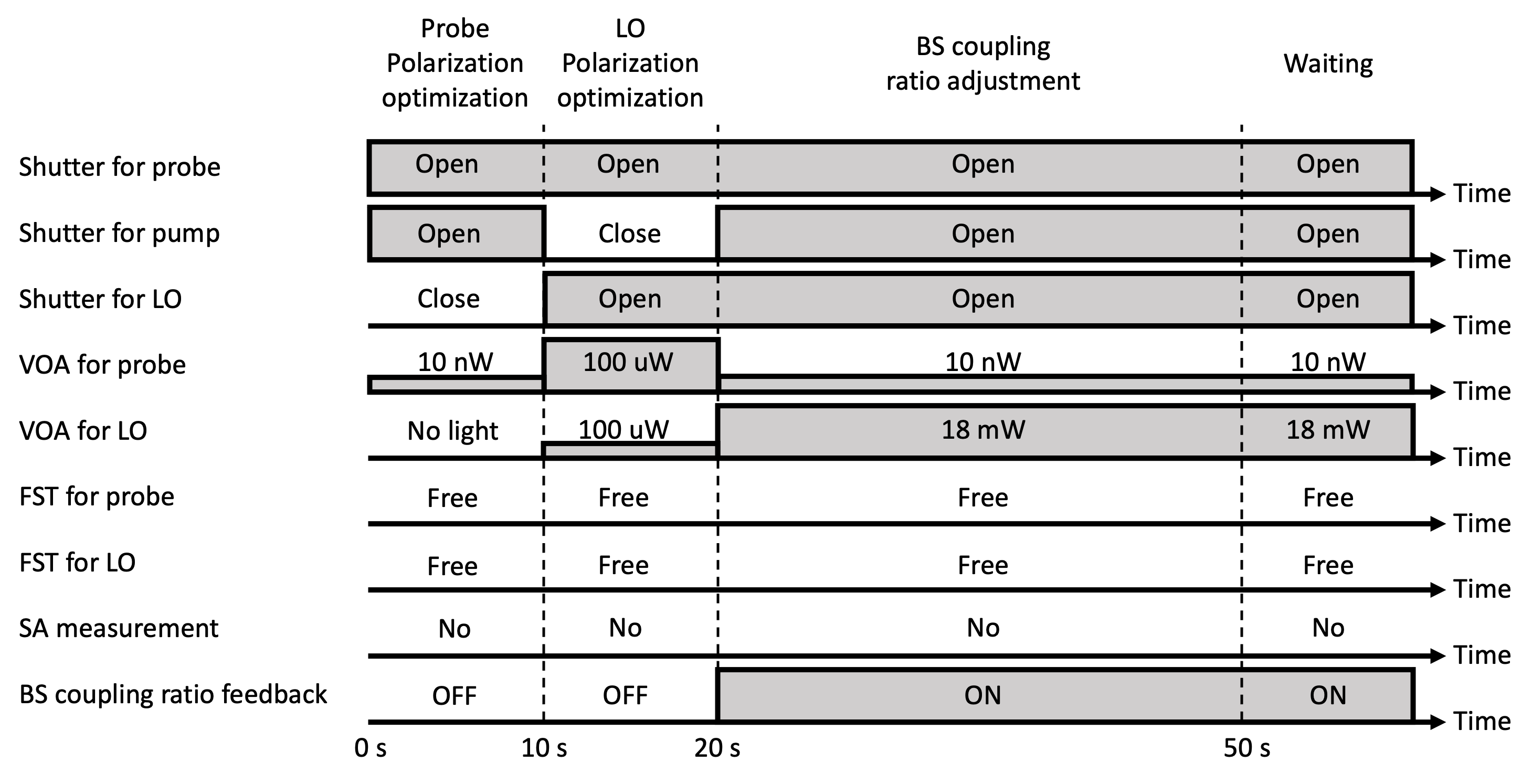}
\caption{Alignment sequence performed every 30 minutes.}
\label{fig12}
\end{figure*}

In this demonstration, automatic alignment is performed once every 30 minutes and squeezed light measurement is performed once every 2 minutes. Alignment is conducted by the sequence in Fig.~\ref{fig12}. Firstly, to match the polarization of the probe light and the crystal axis of the PPLN, only the probe and the pump light go through the shutter, polarization modulation is applied to the probe light, and polarization control is performed using a beat signal  of 400~kHz. Then, to match the polarization of the probe and the LO light, only the probe and the LO light go through the shutter, polarization modulation is applied to the LO light only, and polarization control is performed using a beat signal of 200~kHz. Since the coupling ratio of the beam splitter for homodyne measurement slightly depends on the polarization of the input light, control of the coupling ratio is assigned after polarization control of the probe and the LO light. During the above alignment, the fiber stretcher is fixed at a constant voltage (Unlocked).

\begin{figure*}
\centering
\includegraphics[width=\linewidth]{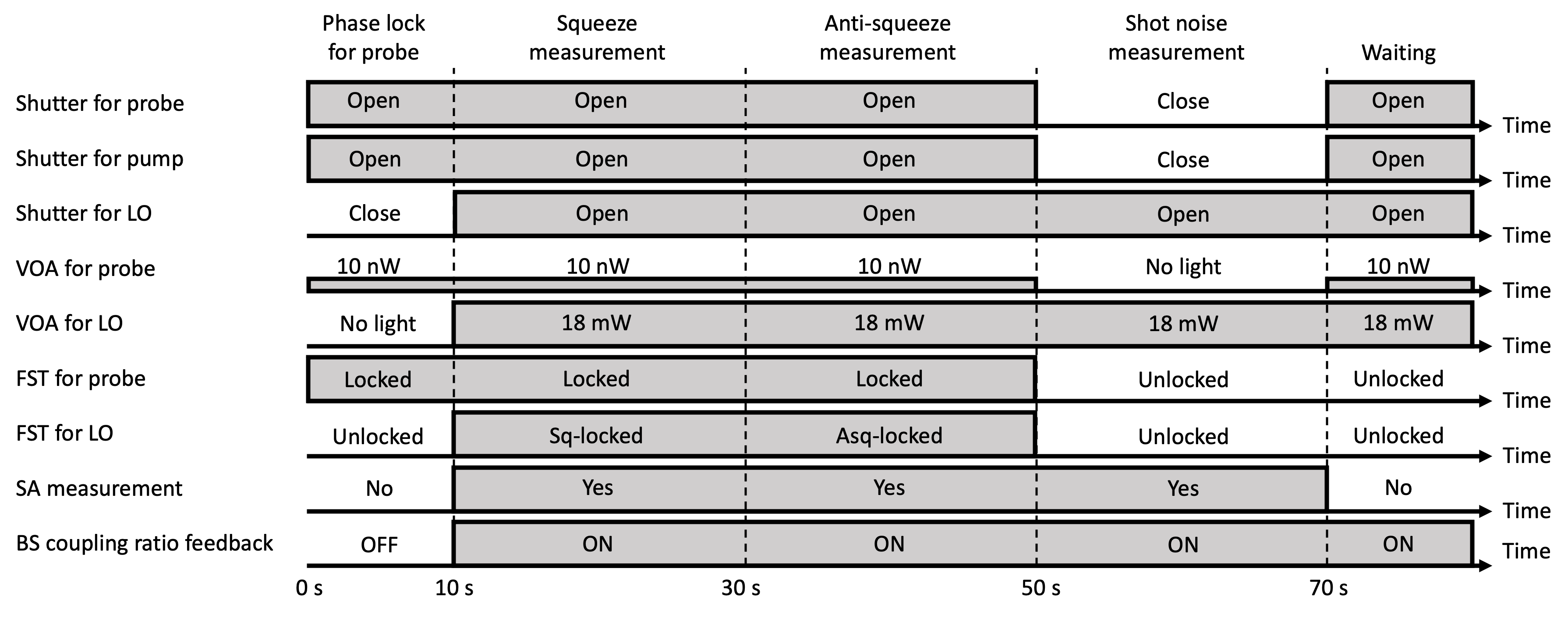}
\caption{Squeezed light measurement sequence performed every 2 minutes.}
\label{fig13}
\end{figure*}

Squeezed light measurements are performed in the sequences shown in Fig.~\ref{fig13}. Firstly, the phase of the probe light is locked with respect to the pump light. When the pump light is input, the temperature of the PPLN inside the OPA increase and the phase drift occurs for a while. Therefore, we wait for 10~seconds so that it settles down. Next, the phase of the LO light is locked with respect to the probe light, and the squeezing and anti-squeezing levels are measured. Finally, for the shot noise level measurement, all phase locks are released and only the LO light is input. During measurement time, control of the coupling ratio of the beam splitter is activated.

\bibliography{library.bib}

\end{document}